\RequirePackage{ifpdf}
\documentclass{PoS}
\pdfoutput = 1
\usepackage{cite}
\usepackage{amsmath}  
\usepackage{amssymb} 
\usepackage{dsfont}
 \usepackage{multirow}
\usepackage{psfrag,scalefnt}
\usepackage{paralist}

\usepackage{color}
\let\normalcolor\relax

\newcommand{\vcb}{|V_{cb}|}

\newcommand{\vub}{|V_{ub}|}

\def\epe{\varepsilon'/\varepsilon}
\newcommand{\tev}{\, {\rm TeV}}

\newcommand{\mev}{\, {\rm MeV}}

\newcommand{\be}{\begin{equation}}
\newcommand{\ee}{\end{equation}}
\newcommand{\bea}{\begin{eqnarray}}
\newcommand{\eea}{\end{eqnarray}}

\newcommand{\ba}{\begin{array}}
\newcommand{\ea}{\end{array}}

\def\kpn{K^+\rightarrow\pi^+\nu\bar\nu}

\def\klpn{K_{L}\rightarrow\pi^0\nu\bar\nu}

\newcommand{\bsi}{B_6^{(1/2)}}
\newcommand{\bei}{B_8^{(3/2)}}

\title{Kaon Theory News}

\ShortTitle{Kaon Theory News}

\author{\speaker{Andrzej J. Buras}%
 \thanks{FLAVOUR(267104)-ERC-110} \\
TUM-IAS, Lichtenbergstr. 2a, D-85748 Garching, Germany \\
Technical University Munich, Physics Department, D-85748 Garching, Germany,\\
E-mail: \email{aburas@ph.tum.de}}

\abstract{During the last fifteen years $B_{s,d}$ decays, $B_{s,d}^0-\bar B^0_{s,d}$ mixings, related mixing induced CP-violating asymmetries $S_{\psi K_S}$ 
and $S_{\psi\phi}$ and CP violation in charm decays provided the dominant 
information about the pattern of flavour violation within the Standard Model (SM) and its extensions. $K$ meson physics was solely represented by the CP-violating parameter $\varepsilon_K$ which by itself could still provide a very powerful
constraint on new physics (NP) models but could not offer by far as much 
information about the pattern of flavour violation in nature as the observables 
in heavy meson decays. We emphasize that in the coming years  $K$ meson physics
will certainly strike back through the measurements of the branching ratios 
of  theoretically very clean decays $\kpn$ and $\klpn$ and improved calculations
of the ratio $\epe$ provided recently by lattice QCD and large $N$ approach. 
Also the improved calculations of the $K^0-\bar K^0$ mass difference $\Delta M_K$ could be of help in this context. We summarize the status of $\kpn$ and 
$\klpn$ within the SM and simplified NP models and update the picture of flavour
 violation in the Littlest Higgs model with T-parity (LHT) taking all available constraints into 
account. But the highlight of this talk are new results on $\epe$ from lattice 
QCD and large $N$ approach that give a strong indication for a new anomaly 
in flavour physics, this time coming from $K$ mesons. Indeed, $\epe$ within 
the SM is found to be significantly below the data of NA48 and KTeV collaborations. Combining lattice results on the non-perturbative parameters $\bsi$ and $\bei$ with the assumption that the SM dynamics dominates the $\Delta I=1/2$ 
rule a $3\sigma$ anomaly in $\epe$ emerges. Moreover, the recently 
derived upper bounds on $\bsi$ and $\bei$ from large $N$ approach 
with $\bsi\le\bei<1$, fully consistent with lattice results, increase the 
confidence that the found anomaly could be an important signal of NP at 
work. Of particular importance are the correlations between $\kpn$, $\klpn$ 
and $\epe$. With future precise measurements of $\kpn$ and $\klpn$ and 
improved calculations of $\epe$ they will surely provide powerful tools for 
 selecting the favourite NP models. We illustrate this with the help of the 
LHT model and simplified models in which NP in $K\to\pi\nu\bar\nu$ 
and $\epe$ is governed by $Z$ with flavour violating couplings or by a heavy 
$Z^\prime$.
 }

\FullConference{The European Physical Society Conference on
                     High Energy Physics\\
                 22-29 July 2015\\
                 Vienna, Austria}

\begin{document}

\section{Overture}
There is no question about that flavour physics will play an important role 
in identifying NP, even if first signs of it would be found at the LHC \cite{Buras:2013ooa,Isidori:2014rba,Buras:2015nta,Fleischer:2015mla}. While the full picture will only 
be gained through the study of flavour violating processes in all meson 
systems and including also lepton flavour violation, breakdown of lepton flavour  universality, electric dipole moments and $(g-2)_{e,\mu}$, in this talk I will
concentrated on $K$ meson physics and the topics already mentioned in the 
abstract. 

This talk is divided into three parts:
\begin{itemize}
\item
Summary of the present status of $\kpn$ and $\klpn$ within the SM followed 
by the recent results obtained in simplified models with flavour violating 
couplings of our $Z$ and of a heavy $Z^\prime$.
\item
Picture of quark flavour observables in the LHT model with T-parity after 
LHC Run 1.
\item
New results on $\epe$ from lattice QCD and large $N$ approach. Here I will 
discuss first of all the emerging anomaly in $\epe$ mentioned in the abstract 
and
the correlation of $\epe$ with $\klpn$ and $\kpn$. 
But, I will also emphasize the compatibility of the results obtained by 
the lattice simulations and the analytic QCD approach to
 $K^0-\bar K^0$ mixing and non-leptonic $K$-meson decays developed in the 1980s
in collaboration with Bardeen and G{\'e}rard. It
is based on the dual representation of QCD as a theory of weakly interacting mesons for large $N$, where $N$ is the number of  colours, as pioneered in 1970s by 't Hooft and Witten.
\end{itemize}

\section{News on $\kpn$ and $\klpn$}
\subsection{Standard Model}
These two very rare decays are exceptional in the flavour physics as 
their branching ratios are know for fixed CKM parameters within an 
uncertainty of $2\%$ which cannot be matched by any other decay to my 
knowledge. Indeed, they are theoretically 
very clean and their branching ratios have been calculated within  the SM including NLO QCD corrections to the top quark contributions 
\cite{Buchalla:1993bv,Misiak:1999yg,Buchalla:1998ba}, 
 NNLO QCD corrections  to the charm contribution in $\kpn$ \cite{Gorbahn:2004my,Buras:2005gr,Buras:2006gb} and  NLO electroweak corrections \cite{Brod:2008ss,Brod:2010hi,Buchalla:1997kz}.
Moreover, extensive calculations of isospin breaking effects and 
non-perturbative effects have been done \cite{Isidori:2005xm,Mescia:2007kn}. 
Therefore, once the CKM parameters $\vcb$, $\vub$ and $\gamma$  will be 
precisely determined in tree-level decays, these two decays will 
offer an excellent test of the SM and constitute a very powerful probe of the NP. Reviews of these two decays can be found in 
\cite{Buras:2004uu,Komatsubara:2012pn,Buras:2013ooa,Blanke:2013goa,Smith:2014mla}. 

It is really exciting that after more than twenty years of waiting \cite{Buchalla:1993bv}, the prospects of measuring the 
branching ratios for these two {\it golden} modes within this decade are very good. Indeed,
the NA62 experiment at CERN is expected to measure the $\kpn$ branching ratio with the precision of $\pm10\%$ \cite{Rinella:2014wfa,Romano:2014xda}, and the KOTO experiment at J-PARC should make a significant progress in measuring the branching ratio for $\klpn$ \cite{Komatsubara:2012pn,Shiomi:2014sfa}.

In \cite{Buras:2015qea} we have reviewed the status of these decays within the SM taking into account all presently available information from other observables and lattice QCD. In calculating the branching ratios for these decays we have first used the tree-level determination of the CKM parameters 
to find
\begin{align}\label{PREDA}
    \mathcal{B}(\kpn) &= \left(8.4 \pm 1.0\right) \times 10^{-11}, \\
    \mathcal{B}(\klpn) &= \left(3.4\pm 0.6\right) \times 10^{-11}.    
\end{align}
This strategy is 
clearly optimal as it allows to predict {\it true} SM values of these branching 
ratios. 

More precise predictions can be obtained by imposing in addition 
the constraints on
CKM parameters from $\Delta F=2$ observables, in 
particular $\varepsilon_K$, $\Delta M_s$, $\Delta M_d$ and mixing induced CP 
asymmetries $S_{\psi K_S}$ and $S_{\psi\phi}$. One finds then  
\begin{align}\label{PREDB}
    \mathcal{B}(\kpn) &= \left(9.1 \pm 0.7\right) \times 10^{-11}, \\
    \mathcal{B}(\klpn) &= \left(3.0\pm 0.3\right) \times 10^{-11}.    
\end{align}
and the expected progress on the determination of weak decay constants and $B_i$ parameters from lattice QCD could certainly reduce the errors by a factor of 
two.

Still, the first strategy of using only tree-level determinations of CKM parameters should be favoured, in particular in the context of NP analyses,  since it allows to determine the CKM matrix elements independently of NP effects. In this 
context, probably the most important results of \cite{Buras:2015qea} are
the following parametric expressions for the branching ratios in terms of the CKM inputs:
\begin{align}
    \mathcal{B}(\kpn) = (8.39 \pm 0.30) \times 10^{-11} \cdot
    \bigg[\frac{\left|V_{cb}\right|}{40.7\times 10^{-3}}\bigg]^{2.8}
    \bigg[\frac{\gamma}{73.2^\circ}\bigg]^{0.74},\label{kplusApprox}
\end{align}
\begin{align}
    \mathcal{B}(\klpn) = (3.36 \pm 0.05) \times 10^{-11} \cdot
    &\bigg[\frac{\left|V_{ub}\right|}{3.88\times 10^{-3}}\bigg]^2
    \bigg[\frac{\left|V_{cb}\right|}{40.7\times 10^{-3}}\bigg]^2
    \bigg[\frac{\sin(\gamma)}{\sin(73.2^\circ)}\bigg]^{2}.
    %&\times\bigg[\frac{\left|V_{us}\right|}{0.2252}\bigg]^{-2}
    %\bigg[\frac{\sin(\gamma)}{\sin(70^\circ)}\bigg]^{2}.
\end{align}
The parametric relation for $\mathcal{B}(\klpn)$ is exact, while for $\mathcal{B}(\kpn)$ it gives an excellent approximation:
for the large ranges $37 \leq |V_{cb}|\times 10^{3} \leq 45$ and $60^\circ \leq \gamma \leq 80^\circ$ it is accurate to 1\% and 0.5\%, respectively. The exposed  errors are non-parametric ones. They originate in the left-over uncertainties
in QCD and electroweak corrections and other small uncertainties. For $\kpn$ the error is larger due to the relevant charm contribution that can be neglected 
for $\klpn$. In the case of $\mathcal{B}(\kpn)$ we have absorbed $|V_{ub}|$ into the non-parametric error due to the weak dependence on it. 

These formulae are useful as they allow easily to monitor the changes in the
values of branching ratios in question, which clearly will still take place before the values on $\vcb$, $\vub$ and $\gamma$ from tree-level decays will be
precisely known. The error budgets can be found in Fig.~1 of \cite{Buras:2015qea}.

Of particular interest are also the correlations between
the branching ratios for $\kpn$ and 
$B_{s}\to\mu^+\mu^-$  in the SM 
\begin{align}
\mathcal{B}(\kpn) = (8.39\pm 0.58)\times 10^{-11} \cdot  \left[\frac{\gamma}{73.2^\circ}\right]^{0.81} 
\left[\frac{\overline{\mathcal{B}}(B_s\to\mu^+\mu^-)}{3.4\times 10^{-9}}\right]^{1.42}\left[\frac{227.7\mev}{F_{B_s}}\right]^{2.84} \label{master1}
\end{align}
and $\kpn$ and $\varepsilon_K$
\begin{align}
\mathcal{B}(\kpn) = (8.39\pm 1.11)\times 10^{-11}\cdot \left[\frac{|\varepsilon_K|}{2.23\times 10^{-3}}\right]^{1.07} 
\left[\frac{\gamma}{73.2^\circ}\right]^{-0.11}\cdot\left[\frac{|V_{ub}|}{3.88\times 10^{-3}}\right]^{-0.95}.\label{master3}
\end{align}

Note that these relations are independent of $\vcb$ and (\ref{master1})
should be of interest in 
the coming years due to  
the measurement of $\kpn$ by NA62, of $B_s\to\mu^+\mu^⁻$ by LHCb and CMS and of $\gamma$ by the LHCb. Moreover the last factor should also be improved by lattice 
QCD. In particular lowering $B_s\to\mu^+\mu^⁻$ through smaller $\vcb$ 
implies $\mathcal{B}(\kpn)$ in the ballpark of $7\cdot 10^{-11}$ and through
 (\ref{master3}) $\varepsilon_K$ below the data unless $\gamma$ and $\vub$ 
are significantly modified.

There are other interesting SM results in \cite{Buras:2015qea}, in particular 
those for the SM, but let us first
turn our attention to our second recent analysis in \cite{Buras:2015yca} which 
was devoted to $K\to\pi\nu\bar\nu$ and $\epe$ in simplified 
NP models. In this part we will only discuss $K\to\pi\nu\bar\nu$ decays. The
results on $\epe$ from this paper will be summarized in the third part of this 
talk.
\subsection{Beyond the Standard Model}
The decays  $\kpn$ and $\klpn$ have been studied over many years in various 
concrete extensions of the SM. A review of the analyses performed until 
August 2007 can be found in \cite{Buras:2004uu}. More recent reviews can be 
found in \cite{Buras:2010wr,Buras:2012ts,Buras:2013ooa,Blanke:2013goa,Smith:2014mla}. Most extensive analyses have been performed in supersymmetric models  \cite{Buras:1997ij,Colangelo:1998pm,Buras:1999da,Buras:2004qb,Crivellin:2011sj}, 
the Littlest Higgs (LH) model without T-parity \cite{Buras:2006wk}, the LH model with T-parity (LHT) \cite{Blanke:2009am,Blanke:2015wba}, Randall-Sundrum models \cite{Blanke:2008yr,Bauer:2009cf}, models with partial compositeness \cite{Straub:2013zca} and  331 models \cite{Buras:2012dp,Buras:2014yna}. All these models 
contain several new parameters related to couplings and masses of new fermions,
 gauge bosons and scalars and the analysis of  $\kpn$ and $\klpn$  requires 
the inclusion of all constraints on couplings and masses of these particles and 
consequently is rather involved. Moreover, the larger number of parameters 
present in these models does not presently allow for clear cut conclusions beyond rough bounds on the size of NP contributions to  $\kpn$ and $\klpn$. 

Therefore, in \cite{Buras:2012jb,Buras:2014sba,Buras:2015yca}, in order to get a better insight into the structure of the possible impact of NP on  $\kpn$ and $\klpn$ decays, and in particular on the correlation between them and other observables, we studied simplified models with flavour violating $Z$ and $Z^\prime$  that contain  small number of parameters. In particular recently, with this idea in mind, we 
analyzed \cite{Buras:2015yca}:
\begin{itemize}
\item
General classes of models based on a $U(3)^3$ flavour symmetry (MFV), illustrating 
them by means of two specific models in which quark flavour violating couplings 
of  $Z$  and of a heavy $Z^\prime$  are consistent with this symmetry.
\item
Models in which the flavour symmetry $U(3)^3$ is reduced to $U(2)^3$, illustrating the results again by means of two simple $Z$ and $Z^\prime$ models.
\item
The $Z$ and $Z^\prime$ models with tree-level FCNCs in which the quark 
couplings are arbitrary 
subject to available constraints from other decays. In particular in this 
case we  included right-handed currents which are absent in MFV and strongly 
suppressed in the simplest $U(2)^3$ models.
\end{itemize}

\begin{figure}[t]
\centering%
\includegraphics[width=0.5\textwidth]{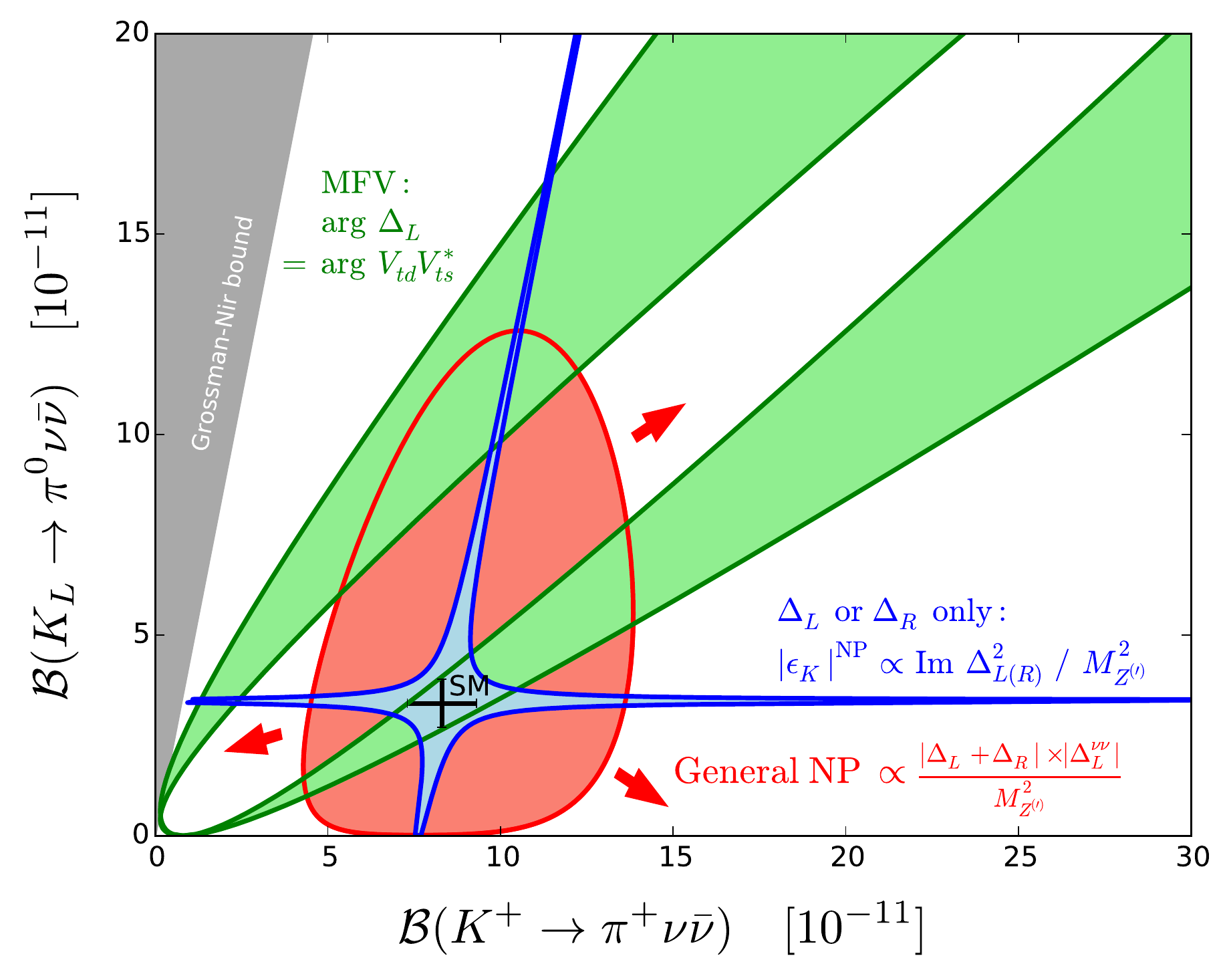}%
\caption{\it Illustrations of common correlations in the $\mathcal{B}(\kpn)$ versus $\mathcal{B}(\klpn)$ plane. The expanding red region illustrates the lack of correlation for models with general LH and RH NP couplings. The green region shows the correlation present in models obeying CMFV. The blue region shows the correlation induced by the constraint from $\varepsilon_K$ if only LH or RH couplings  are present. From \cite{Buras:2015yca}. \label{fig:illustrateEpsK}}
\end{figure}

The main results of this study, postponing the results for $\epe$ to the third 
part of this talk, are as follows:
\begin{itemize}
\item
There is a hierarchy in the size of possible NP effects in $K\to\pi\nu\bar\nu$ 
mediated by tree-level $Z$ and $Z^\prime$ exchanges. 
They are smallest in MFV models, larger in $U(2)^3$ models and significantly larger in the case of new sources of flavour and CP violation beyond these two CKM-like frameworks.
\item
In MFV models NP effects in $K\to\pi\nu\bar\nu$ above $30\%$ at the level of the branching are rather unlikely. Moreover, there is as strong correlation between
the branching ratios for these two decays. An important constraint on the 
 size of NP comes from the data on $B_s\to\mu^+\mu^-$ decay.
\item
The latter constraint is absent in $U(2)^3$ models allowing for larger NP
effects.
\item
In $Z^\prime$ models with MFV the present $B_d\to K (K^*) \mu^+\mu^-$ anomalies favour the enhancement of $\kpn$ and $\klpn$.
$\Delta F = 2$ observables however put significant constraints on this possibility.
\item
Beyond MFV and $U(2)^3$ models the correlation between two branching ratios 
and the size of possible NP effects 
depends on the presence or absence of $\varepsilon_K$ constraint \cite{Blanke:2009pq}, the impact 
of $K_L\to \mu^+\mu^-$ that depends on whether the flavour violating couplings
are left-handed or right-handed \cite{Blanke:2008yr}, and in the case of $\klpn$, as we will see below, 
on the $\epe$ constraint \cite{Buras:1998ed}. Still in the case of $Z$ FCNCs enhancements by factors of 2-3 over the SM expectations are still possible.
Due to the absence of correlations between $K\to\pi\nu\bar\nu$ and $\epe$ 
in general $Z^\prime$ models, the size of NP contributions in these decays could be in principle even larger. Then, as demonstrated
in \cite{Buras:2014zga}, $\kpn$ and $\klpn$ can probe energy scales as large as $1000\tev$ in the presence of general 
flavour-violating couplings.
\item
Finally, the future measurement of $\mathcal{B}(\klpn)$ and its correlation 
with $\mathcal{B}(\kpn)$ will significantly facilitate the distinction between various models. See Fig.~\ref{fig:illustrateEpsK} and related text in \cite{Blanke:2009pq,Buras:2015yca}.
\end{itemize}

\section{LHT Model Facing New Flavour Data}
The Littlest Higgs Model with T-parity (LHT) 
belongs to the simplest NP scenarios with new sources of flavour 
and CP violation. The latter originate in the interactions of ordinary quarks 
and leptons with heavy mirror quarks and leptons that are mediated 
by new  heavy gauge bosons. 
Also a heavy fermionic top partner is present in this model which communicates with 
the SM fermions by means of standard $W^\pm$ and $Z^0$ gauge bosons.
In \cite{Blanke:2015wba} we have presented a new analysis of quark  observables in the LHT model in view of 
the oncoming flavour precision era. To this end we used all available information on the CKM parameters, lattice QCD input and experimental data on quark flavour observables and corresponding theoretical calculations, 
taking into account new lower bounds on the symmetry breaking scale and the mirror quark masses from the LHC.  Our main findings are as follows:
\begin{itemize}
\item
The LHT model agrees well with the data on $\Delta F=2$ observables and is 
capable of removing some slight tensions between the SM predictions and the data.
\item
The most interesting departures from SM predictions can be found for $\kpn$ and 
$\klpn$ decays, when only constraints from  $\Delta F=2$ observables are taken 
into account. {An enhancement} of the branching ratio for $\kpn$  by a factor of two relative to the SM prediction quoted above is {still} possible. An even 
larger enhancement in the case of $\klpn$ is allowed. But as we will 
discuss soon the recent news on $\epe$ in the SM 
appears to exclude this possibility at present. Rather a suppression of  $\klpn$ is required to fit the data on $\epe$. On the other hand 
 no significant shifts of $\kpn$ with respect to SM are then allowed. 
\item
NP effects in rare $B_{s,d}$ decays are significantly smaller than in rare $K$ 
decays.  Still they can amount to up to a factor of 2 in the $b\to d$ system and to about $50\%$ of the SM branching ratios in $b\to s$ transitions, like $\mathcal{B}(B_s\to\mu^+\mu^-)$ and $B\to K^{(*)}\nu\bar\nu$. 
\item
{More interestingly} the pattern of departures from SM expectations for $B_{s,d}$ decays predicted by the LHT model 
disagrees with the present data. $\mathcal{B}(B_s\to\mu^+\mu^-)$ is favoured 
by this model to be enhanced rather than suppressed as indicated by the data, and the simultaneous enhancement of  $\mathcal{B}(B_d\to\mu^+\mu^-)$ cannot be explained.
More importantly the LHT model fails to reproduce the 
$B_d\to K^{(*)}\ell^+\ell^-$ and $B^+\to K^+\ell^+\ell^-$anomalies observed by the LHCb experiment.
\end{itemize}

The future of the LHT model depends crucially on the improved experimental 
values of  $\mathcal{B}(B_{s,d}\to\mu^+\mu^-)$ and on the future of the 
$B_d\to K^{(*)} \ell^+\ell^-$ anomalies. If these anomalies will be confirmed 
by future more accurate data and theory predictions, then the LHT model is
 not the NP realised in nature. For this model to survive the flavour tests in the quark sector, the anomalies in question have to disappear. Then the LHT 
model will have to face $\epe$ as we will see soon. For other news on LHT, in 
particular if no LHT particle will be found at the LHC, we refer to 
 \cite{Blanke:2015wba}.

\section{$\epe$ Striking Back}
\subsection{Standard Model}
One of the stars of flavour physics in the 1990s  was the ratio
$\epe$  that measures the size of the direct CP violation in $K_L\to\pi\pi$ 
relative to the indirect CP violation described by $\varepsilon_K$. On the 
experimental side after heroic efforts on both sides of Atlantic
the world average from NA48 \cite{Batley:2002gn} and KTeV
\cite{AlaviHarati:2002ye,Abouzaid:2010ny} collaborations reads
\be\label{EXP}
(\epe)_\text{exp}=(16.6\pm 2.3)\times 10^{-4} \,.
\ee

On the theory side a long-standing challenge in  making predictions for $\epe$ within the SM and its extensions has been the strong interplay of QCD penguin contributions and electroweak penguin contributions to this ratio. In the SM, QCD penguins
give a positive contribution and electroweak penguins a negative one. In order
to obtain a useful prediction for $\epe$, the relevant contributions of the QCD
penguin and electroweak penguin operators must be know accurately.
Reviews on $\epe$ can be found in 
\cite{Bertolini:1998vd,Buras:2003zz,Pich:2004ee,Cirigliano:2011ny,Bertolini:2012pu}.

As far as short-distance contributions (Wilson coefficients of QCD and 
electroweak penguin operators) are concerned, they have been known already
for more than twenty years at the NLO level
\cite{Buras:1991jm,Buras:1992tc,Buras:1992zv,Ciuchini:1992tj,Buras:1993dy,Ciuchini:1993vr}
and present technology could extend them to the NNLO level if necessary. First
steps in this direction have been taken in \cite{Buras:1999st,Gorbahn:2004my,Brod:2010mj}.

The situation with hadronic matrix elements is another story and even if 
significant progress on their evaluation has been made  over the last 25 years,
the present status is clearly not satisfactory. In order to describe the 
problem in explicit terms let me write down the formula for $\epe$ recently 
presented in \cite{Buras:2015yba}
\begin{equation}
\frac{\varepsilon'}{\varepsilon} = 10^{-4} \biggl[
\frac{{\rm Im}\lambda_{\rm t}}{1.4\cdot 10^{-4}}\biggr]\!\left[\,a\,
\big(1-\hat\Omega_{\rm eff}\big) \big(-4.1(8) + 24.7\,\bsi\big) + 1.2(1) -
10.4\,\bei \,\right]\,.
\label{AN2015}
\end{equation}
This formula has been obtained by assuming that the real parts of the $K\to\pi\pi$ isospin amplitudes $A_0$ and $A_2$, which exhibit the $\Delta I=1/2$ rule,
are fully described by SM dynamics. Their experimental values are used to
determine to a very good approximation hadronic matrix elements of all
$(V-A)\otimes (V-A)$ operators  \cite{Buras:1993dy}. In this manner the main uncertainties in $\epe$ reside in the parameters $\bsi$ and $\bei$ which parametrize the hadronic matrix elements of the $(V-A)\otimes (V+A)$ QCD penguin and
electroweak penguin operators, $Q_6$ and $Q_8$, respectively. The first and the third  term in (\ref{AN2015}) summarize respectively the contributions of $(V-A)\otimes (V-A)$ QCD and electroweak penguin operators that have been extracted 
using the experimental data on the real parts of $A_0$ and $A_2$. 

The parameters  $a$ and $\hat\Omega_{\rm eff}$ summarize  isospin breaking corrections and include  strong isospin
violation $(m_u\neq m_d)$, the correction to the isospin limit coming from
$\Delta I=5/2$ transitions and  electromagnetic corrections and can be 
extracted from \cite{Cirigliano:2003nn,Cirigliano:2003gt}. They are given as follows \cite{Buras:2015yba}
\be\label{OM+}
 a=1.017, \qquad
\hat\Omega_{\rm eff} = (14.8\pm 8.0)\times 10^{-2}\,.
\ee

Recently significant progress on the values of $\bsi$ and $\bei$ has been 
made by the RBC-UKQCD collaboration, who presented their results on 
the relevant hadronic matrix elements of the operator $Q_6$ \cite{Bai:2015nea}
and $Q_8$ \cite{Blum:2015ywa}. These results imply the following values 
for $\bsi$ and $\bei$ \cite{Buras:2015yba,Buras:2015qea}
\be\label{Lbsi}
\bsi=0.57\pm 0.19\,, \qquad \bei= 0.76\pm 0.05\,, \qquad (\mbox{RBC-UKQCD})
\ee
to be compared with their values in the strict large $N$ limit of QCD
\cite{Buras:1985yx,Bardeen:1986vp,Buras:1987wc}
\be\label{LN}
\bsi=\bei=1, \qquad {\rm (large~N~Limit)}\,.
\ee

The low value of $\bsi$ in (\ref{Lbsi}) is at first sight surprising and as it is based on a
numerical simulation one could wonder whether it is the result of a statistical
fluctuation. But the very recent analysis in the large-$N$ approach in
\cite{Buras:2015xba} gives strong support to the values in  (\ref{Lbsi}).
In fact, in this analytic approach one can demonstrate explicitly the
suppression of both $\bsi$ and $\bei$ below their large-$N$ limit 
and  derive a conservative upper bound on both $\bsi$ and $\bei$ which reads
\cite{Buras:2015xba}
\be\label{NBOUND}
\bsi\le \bei < 1 \, \qquad (\mbox{\rm large-}N).
\ee
While one finds $B_8^{(3/2)}(m_c)=0.80\pm 0.10$, the result for $\bsi$ is less
precise but there is a strong indication that $\bsi < \bei$ in agreement with
(\ref{Lbsi}). For further details, see \cite{Buras:2015xba}.

Combining the lattice results in (\ref{Lbsi})  with (\ref{AN2015}) a detailed 
numerical analysis in \cite{Buras:2015yba} gives
\be\label{LBGJJ}
   \epe = (1.9 \pm 4.5) \times 10^{-4} \,,
\ee
roughly $3\sigma$ away from the experimental value in (\ref{EXP}).

But even discarding the lattice results, varying all input parameters, we find
at the bound $\bsi=\bei=1$,
\be\label{BoundBGJJ}
(\epe)_\text{SM}= (8.6\pm 3.2) \times 10^{-4} \,,
\ee
still $2\,\sigma$  below the experimental data. We consider this bound
conservative since employing the lattice value for $\bei$ in (\ref{Lbsi}) and
$\bsi=\bei=0.76$, instead of (\ref{BoundBGJJ}), one obtains
$(6.0\pm 2.4)\times 10^{-4}$. We observe then that even for these values of $\bsi$ and $\bei$ the SM predictions
for $\epe$ are significantly below the data. This is an important result as it 
shows that even if the value of $\bsi$ from lattice calculations would move up 
in the future, the SM would face difficulty in reproducing the data provided
the large-$N$ bound in  (\ref{NBOUND}) is respected. Thus it appears that 
we have a new anomaly in flavour physics, this time coming from the $K$ meson
sector. 

In contrast to our analysis the RBC-UKQCD lattice collaboration \cite{Bai:2015nea} does not include isospin breaking effects and calculates all hadronic matrix
elements directly, that is not imposing the $\Delta I=1/2$ rule. It is then 
not surprizing that with their values in (\ref{Lbsi}) they get much 
less precise result for $\epe$
\be\label{RBC}
(\epe)_\text{SM} = (1.4 \pm 7.0)\times 10^{-4} \,,
\ee
but also this result indicates that SM has some problems in 
reproducing the data.

As the bound in (\ref{NBOUND}) plays a significant role in the conclusion that
NP could be at work in $\epe$, let me ask sceptical readers to have a look 
at \cite{Buras:2015xba,Buras:2014apa} where other successes of the large N 
approach \cite{Buras:1985yx,Bardeen:1986vp,Bardeen:1986uz,Bardeen:1986vz,Bardeen:1987vg,Buras:2014maa} are summarized. In particular those related to the 
$\Delta I=1/2$ rule and the $\hat B_K$ parameter entering $\varepsilon_K$ that 
after almost three decades are supported by lattice QCD. Therefore, I strongly
believe that future more precise lattice calculations of $\bsi$ and $\bei$ will confirm the bound in (\ref{NBOUND}) implying that indeed NP contributes 
significantly to $\epe$ unless the error in the  experimental value in (\ref{EXP}) has been underestimated.

The  present situation with $\epe$ reminds us  the story of $(g-2)_\mu$, where after fifteen years of the Brookhaven result we are not fully confident, 
whether NP is at work here. New experiment at Fermilab should be able to 
clarify this situation at the end of this decade. Assuming that other lattice 
groups will join RBC-UKQCD efforts in calculating $\bsi$ and $\bei$, we 
should be able to decide around 2020 whether a new experiment on $\epe$ is 
required. To this end also the roles of final state interactions (FSI) and of NNLO QCD corrections in  $\epe$ have to be better understood. While in 
\cite{Pallante:1999qf} the enhancement of $\epe$ through FSI is claimed (see 
however \cite{Buras:2000kx}), the NNLO corrections to electroweak penguins 
contributions to $\epe$ suppress this ratio \cite{Buras:1999st}. The NNLO QCD
corrections to QCD penguins are unknown at present.

\begin{figure}
\centering
\includegraphics[width=.40\textwidth]{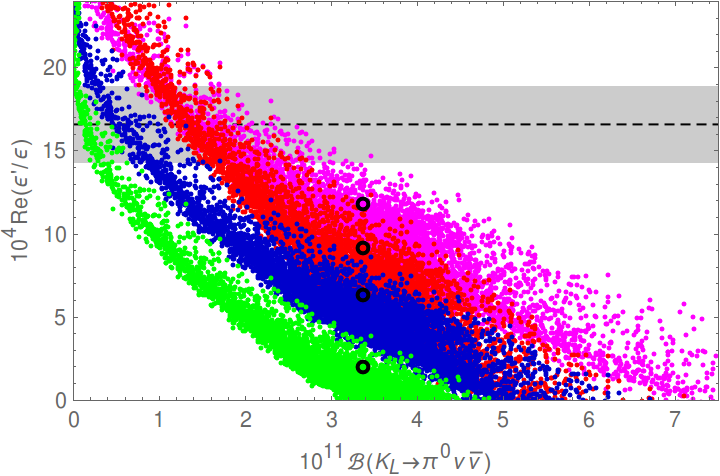}\hskip1cm
\includegraphics[width=.40\textwidth]{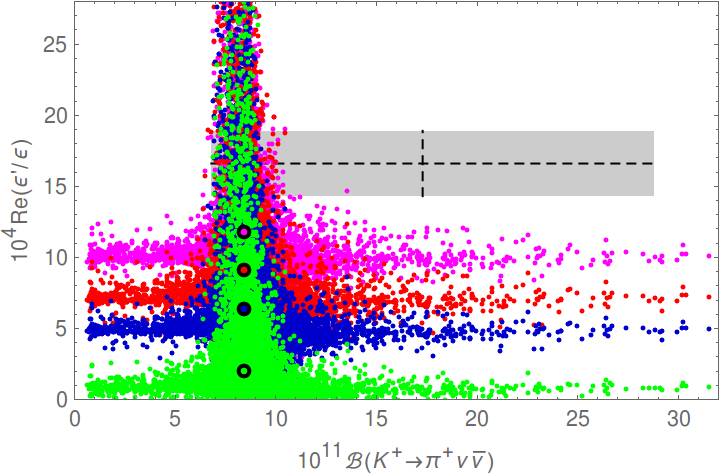}
\caption{\label{fig:epe}\it Correlation between $\mathcal{B}(\klpn)$ and $\epe$ (left panel) and $\mathcal{B}(\kpn)$ and $\epe$ (right panel) in the LHT model for $f=1\tev$ for different values of $(\bsi,\bei)$: $(1.0,1.0)$ (red), $(0.76, 0.76)$ (blue), $(0.57,0.76)$ (green), $(1.0,0.76)$ (magenta). The black dots show the corresponding central SM values. The experimental $1\sigma$ range for $\epe$ is displayed by the grey band. From  \cite{Blanke:2015wba}.}
\end{figure}

\subsection{Beyond the Standard Model}
In most extensions of the SM the enhancement of the branching ratio for 
 $\klpn$ through NP usually implies the suppression of $\epe$ and enhancement of $\epe$ implies suppression of $\klpn$. 
This is related
to the fact that there is a strong correlation between {\it negative} electroweak penguin contribution to $\epe$ and the branching ratios for $\klpn$. 
The correlation with $\kpn$, which is CP-conserving, is weaker but it exists in 
specific models.

 In the left panel of Fig.~\ref{fig:epe} we show the correlation between $\klpn$ and $\epe$ in the LHT model \cite{Blanke:2015wba}. In the right panel the analogous correlation between $\mathcal{B}(K^+\to\pi^+\nu\bar\nu)$ and $\epe$ is shown. Different colours 
correspond to different choices of $(\bsi,\bei)$. We observe that LHT model can 
reproduce the experimental value of $\epe$ provided $\klpn$ decay is strongly 
suppressed with respect to its SM value, while NP effects in $\kpn$ are small. 
NP effects are governed here by  electroweak penguins.

\begin{figure}
\centering%
\includegraphics[width=0.32\textwidth]{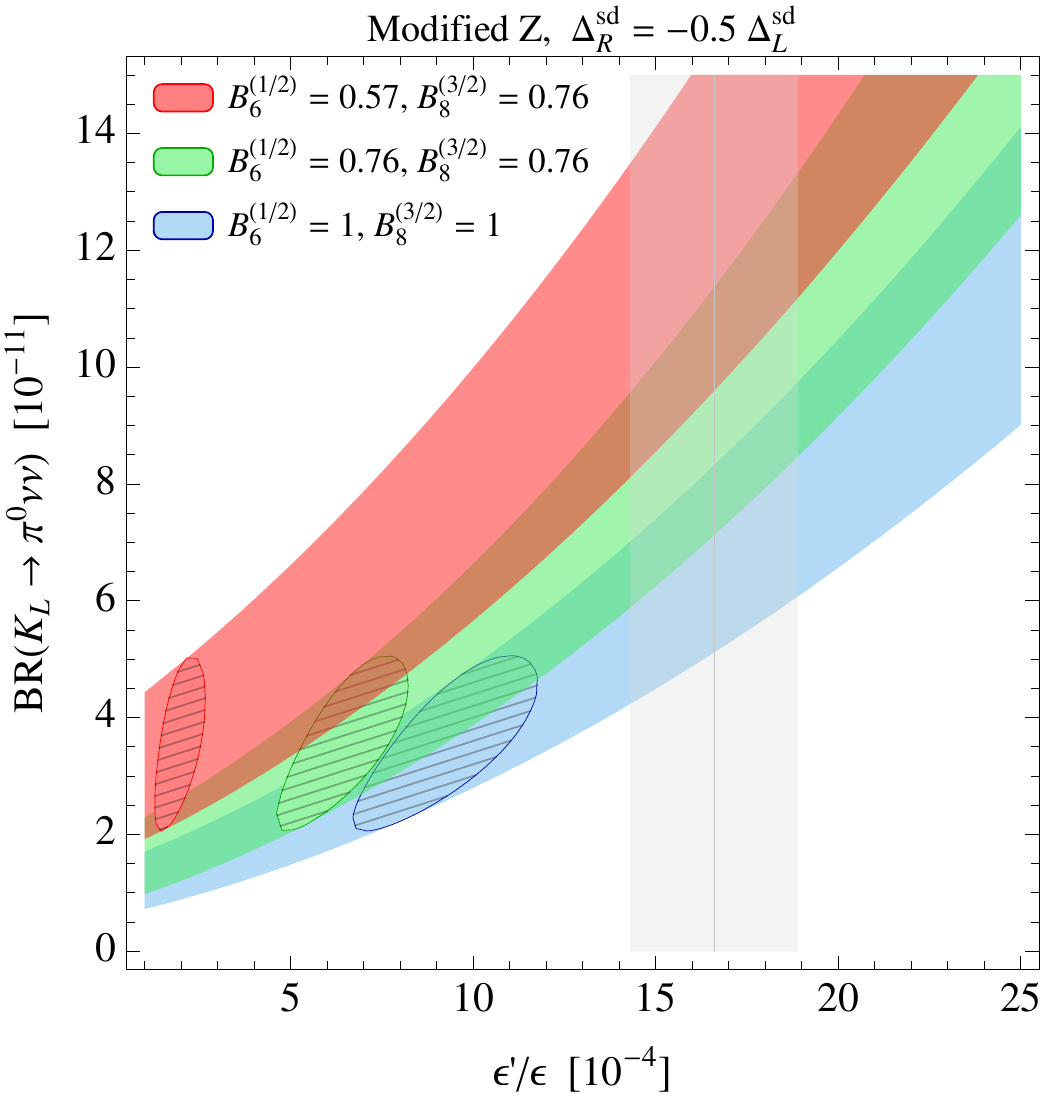}\hfill%
\includegraphics[width=0.32\textwidth]{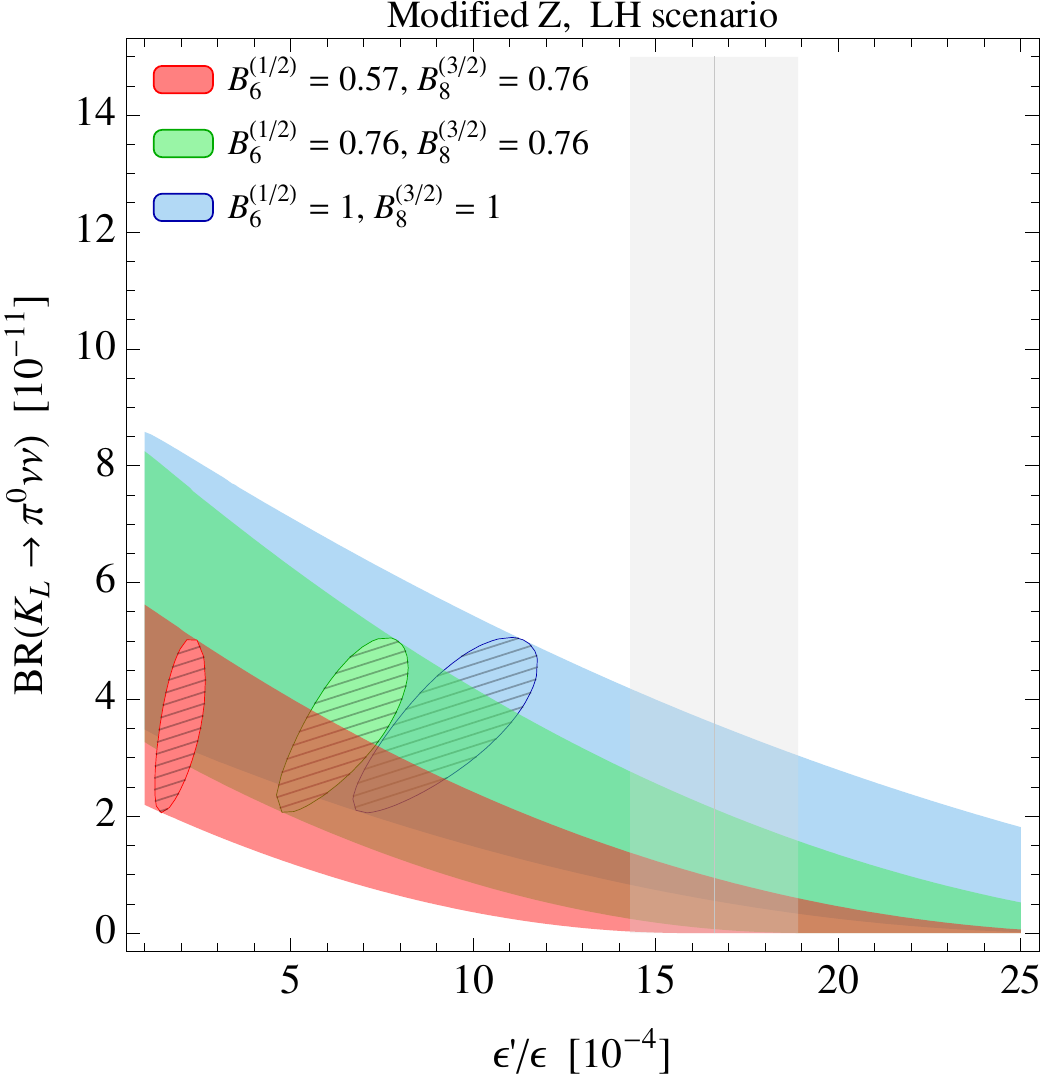}\hfill%
\includegraphics[width=0.32\textwidth]{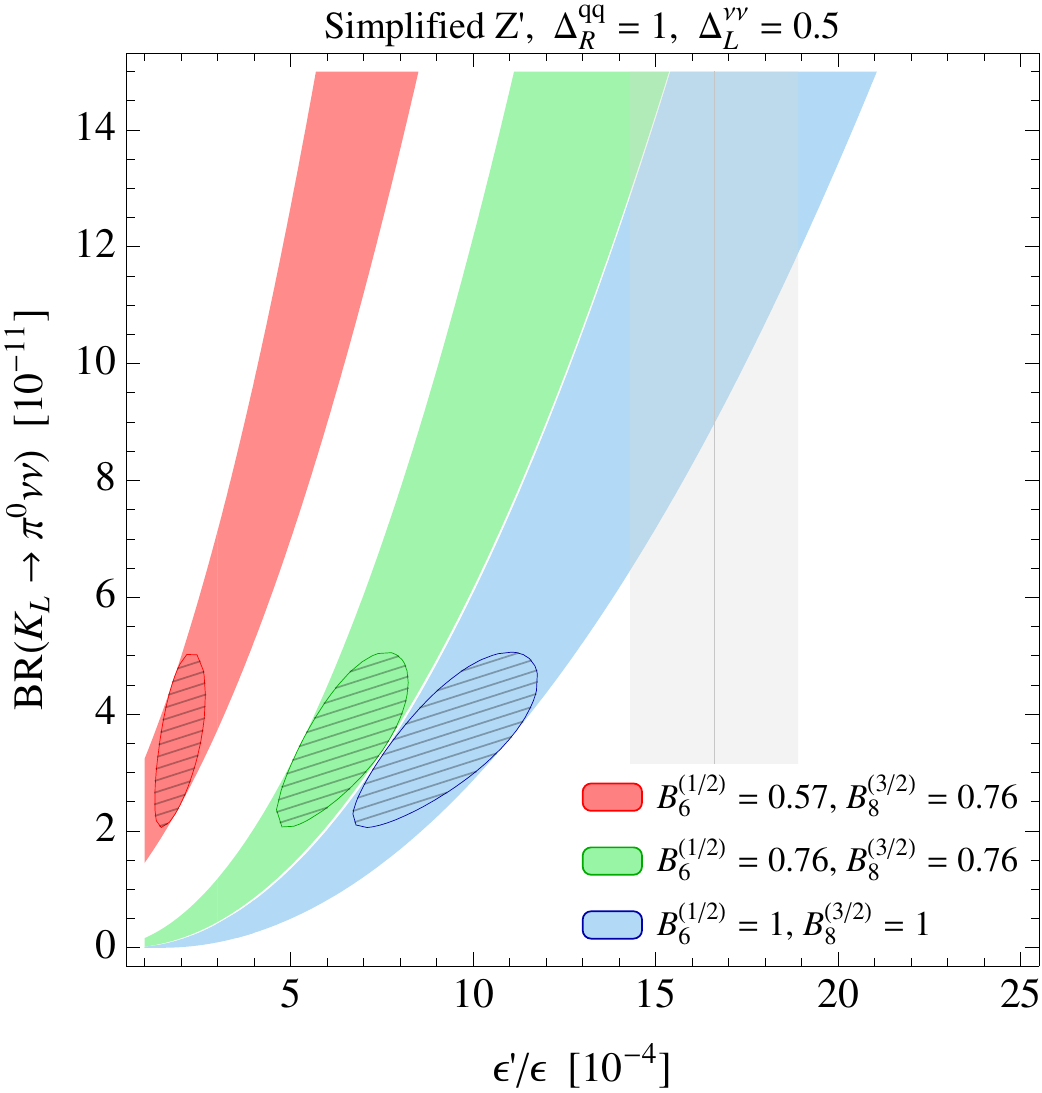}
\caption{\it 95\% C.L. allowed regions for $\epe$ and $\klpn$. Left: model with flavour-changing Z boson couplings $\Delta_R^{sd} = -0.5 \Delta_L^{sd}$. Center: modified Z, LH scenario $\Delta_R^{sd} = 0$. Right: 5 TeV Z' with $\Delta_R^{qq} = 1$ and $\Delta_L^{\nu\nu} = 0.5$. The plots are for $B_6 = 1$ (blue), $B_6 = 0.76$ (green), and $B_6 = 0.57$ (red). The hatched regions are the SM predictions at $2\sigma$. The gray band shows the experimental result for $\epe$. From 
\cite{Buras:2015yca}. 
\label{ZepsilonKL}}
\end{figure}

However, as demonstrated in \cite{Buras:2015yca} one can construct 
 simplified models in which in fact 
$\epe$ and the branching ratio for $\klpn$ can be simultaneously 
enhanced with respect to their SM values. 
In the left panel of Fig.~\ref{ZepsilonKL} we show the correlation between $\epe$ and $\klpn$ in a model in which $Z$ has both the left-handed and right-handed flavour-violating couplings $\Delta_{L,R}^{sd}(Z) \not=0$, and compare it with the opposite correlation that is present in the LH scenario in which 
the right-handed couplings vanish (central panel). The latter result is 
then similar to the one in the LHT model. 

It turns out then that in the case of $Z$ the simultaneous enhancement of $\epe$ 
and $\klpn$ requires the existence of flavour-violating right-handed $Z$ 
couplings. As shown in the right panel of Fig.~\ref{ZepsilonKL} this is not
required in the case of $Z^\prime$.  A tree-level exchange 
of $Z^\prime$ with left-handed flavour violating quark couplings and flavour universal structure of diagonal RH quark couplings contributes to $\epe$ 
 dominantly through the QCD penguin operator $Q_6$ and is capable of enhancing 
$\epe$ and $\klpn$ simultaneously \cite{Buras:2014sba,Buras:2015yca}.

\section{Summary}
$K$ meson flavour physics will surely strike back through the 
measurements of the branching ratios $\kpn$ and $\klpn$ and the improved 
theory of $\epe$. As pointed out already in \cite{Buras:1994ec} precise measurements of both $K\to\pi\nu\bar\nu$ branching ratios would offer the determination 
of the unitarity triangle which could be compared with the one extracted these
days dominantly from $B$ physics. In particular as demonstrated in 
 \cite{Buchalla:1994tr,Buras:2001af} within the SM and models with MFV, 
rather precise determination of $\sin 2\beta$ without the usual QCD 
penguin pollution and almost independently of $\vcb$ can be obtained. Analytic 
expressions for the parameters $\bar\varrho$ and $\bar\eta$ in terms of 
the branching ratios for $\kpn$ and $\klpn$ can be found in these papers and 
numerical analyses have been presented by us in several papers since then.
Recently also the authors of \cite{Lehner:2015jga} have shown that the inclusion of 
$\varepsilon_K$ and $\epe$ in addition to  $\kpn$ and $\klpn$ could help 
to determine such K-triangle. 

Even if I happened to start such a game in \cite{Buras:1994ec}, when we expected
that the branching ratios for $\kpn$ and $\klpn$ will be measured at Brookhaven 
and Fermilab around the year 2000, I do not think that this strategy is a very 
efficient way to search for NP. With already rather precise values of CKM 
parameters, obtained dominantly from $B$ physics experiments, it appears to me 
that it is more straightforward to identify NP by simply comparing precise SM 
predictions for both branching ratios and $\epe$ with the future precise data. In this 
manner not only departures of SM predictions from data can be hopefully identified but the pattern of these deviations will give us some hints what kind of NP
could be responsible for these deviations.

A similar comment applies to the very old idea pioneered in \cite{Gaillard:1974hs} in the case of the determination of the charm quark mass from $K^0-\bar K^0$ mixing and the top quark 
mass from $B_d-\bar B_d$ mixing \cite{Ellis:1977uk} that has been extended 
to rare decays, in particular $\kpn$ and 
$\klpn$, in \cite{Buras:1992uf}. It has recently  been reconsidered in the 
context of general indirect determinations of $m_t$ in \cite{Giudice:2015toa}. 
Even if such studies have some virtues, similar to the determination of 
$\vcb$  from rare Kaon decays \cite{Buras:1994rj}, they necessarily 
assume that NP does not contribute to the processes used for such determinations
 at a measurable level and we should really hope that the nature will tell us soon that this assumption is wrong.

 In any case I am convinced that $\kpn$, $\klpn$ and $\epe$  will in the coming ten years contribute in 
a very important manner to our understanding of the short distance dynamics, 
in particular to the one beyond the reach of the LHC \cite{Buras:2014zga}.

{\bf  Acknowledgements}\\
I thank Dario Buttazzo, Fulvia De Fazio, 
Jennifer Girrbach-Noe and Rob Knegjens for exciting time we spent together 
analyzing the decays $\kpn$ and $\klpn$ within the SM and various NP models.
Particular thanks go to Monika Blanke and Stefan Recksiegel for returning with 
me after six years to the LHT model. Very enjoyable was also the great 
collaboration with Jean-Marc G{\'e}rard, Martin Gorbahn, Sebastian J{\"a}ger and  Matthias Jamin on $\epe$.
Finally, I would like to thank the organizers of the flavour session for inviting me to present these results.
The research presented in this report was dominantly financed and done in the context of the ERC Advanced Grant project ``FLAVOUR'' (267104).  It was also partially supported by the 
DFG cluster of excellence ``Origin and Structure of the Universe''.

\bibliographystyle{JHEP}
\bibliography{allrefs}

\providecommand{\href}[2]{#2}\begingroup\raggedright\begin{thebibliography}{10}

\bibitem{Buras:2013ooa}
A.~J. Buras and J.~Girrbach, {\it {Towards the Identification of New Physics
  through Quark Flavour Violating Processes}},  {\em Rept.Prog.Phys.} {\bf 77}
  (2014) 086201, [\href{http://arxiv.org/abs/1306.3775}{{\tt
  arXiv:1306.3775}}].

\bibitem{Isidori:2014rba}
G.~Isidori and F.~Teubert, {\it {Status of indirect searches for New Physics
  with heavy flavour decays after the initial LHC run}},  {\em Eur.Phys.J.Plus}
  {\bf 129} (2014) 40, [\href{http://arxiv.org/abs/1402.2844}{{\tt
  arXiv:1402.2844}}].

\bibitem{Buras:2015nta}
A.~J. Buras, {\it {Flavour Expedition to the Zeptouniverse}},  {\em PoS} {\bf
  FWNP} (2015) 003, [\href{http://arxiv.org/abs/1505.00618}{{\tt
  arXiv:1505.00618}}].

\bibitem{Fleischer:2015mla}
R.~Fleischer, {\it {Theoretical Prospects for B Physics}},
  \href{http://arxiv.org/abs/1509.00601}{{\tt arXiv:1509.00601}}.

\bibitem{Buchalla:1993bv}
G.~Buchalla and A.~J. Buras, {\it Qcd corrections to rare $k$ and $b$ decays
  for arbitrary top quark mass},  {\em Nucl. Phys.} {\bf B400} (1993) 225--239.

\bibitem{Misiak:1999yg}
M.~Misiak and J.~Urban, {\it {QCD corrections to FCNC decays mediated by Z
  penguins and W boxes}},  {\em Phys.Lett.} {\bf B451} (1999) 161--169,
  [\href{http://arxiv.org/abs/hep-ph/9901278}{{\tt hep-ph/9901278}}].

\bibitem{Buchalla:1998ba}
G.~Buchalla and A.~J. Buras, {\it {The rare decays $K\to\pi \nu\bar\nu$, $B\to
  X \nu\bar\nu$ and $B\to \ell^+\ell^-$: An Update}},  {\em Nucl.Phys.} {\bf
  B548} (1999) 309--327, [\href{http://arxiv.org/abs/hep-ph/9901288}{{\tt
  hep-ph/9901288}}].

\bibitem{Gorbahn:2004my}
M.~Gorbahn and U.~Haisch, {\it {Effective Hamiltonian for non-leptonic $|\Delta
  F| = 1$ decays at NNLO in QCD}},  {\em Nucl.Phys.} {\bf B713} (2005)
  291--332, [\href{http://arxiv.org/abs/hep-ph/0411071}{{\tt hep-ph/0411071}}].

\bibitem{Buras:2005gr}
A.~J. Buras, M.~Gorbahn, U.~Haisch, and U.~Nierste, {\it {The rare decay $K^+
  \to \pi^+ \nu \bar\nu$ at the next-to-next-to-leading order in QCD}},  {\em
  Phys. Rev. Lett.} {\bf 95} (2005) 261805,
  [\href{http://arxiv.org/abs/hep-ph/0508165}{{\tt hep-ph/0508165}}].

\bibitem{Buras:2006gb}
A.~J. Buras, M.~Gorbahn, U.~Haisch, and U.~Nierste, {\it {Charm quark
  contribution to $K^+ \to \pi^+ \nu \bar\nu$ at next-to-next-to-leading
  order}},  {\em JHEP} {\bf 11} (2006) 002,
  [\href{http://arxiv.org/abs/hep-ph/0603079}{{\tt hep-ph/0603079}}].

\bibitem{Brod:2008ss}
J.~Brod and M.~Gorbahn, {\it {Electroweak Corrections to the Charm Quark
  Contribution to $K^+ \to \pi^+ \nu \bar\nu$}},  {\em Phys. Rev.} {\bf D78}
  (2008) 034006, [\href{http://arxiv.org/abs/0805.4119}{{\tt
  arXiv:0805.4119}}].

\bibitem{Brod:2010hi}
J.~Brod, M.~Gorbahn, and E.~Stamou, {\it {Two-Loop Electroweak Corrections for
  the $K \to \pi \nu \bar{nu}$ Decays}},  {\em Phys.Rev.} {\bf D83} (2011)
  034030, [\href{http://arxiv.org/abs/1009.0947}{{\tt arXiv:1009.0947}}].

\bibitem{Buchalla:1997kz}
G.~Buchalla and A.~J. Buras, {\it Two-loop large-$m_t$ electroweak corrections
  to $k \to\pi\nu\bar\nu$ for arbitrary higgs boson mass},  {\em Phys. Rev.}
  {\bf D57} (1998) 216--223, [\href{http://arxiv.org/abs/hep-ph/9707243}{{\tt
  hep-ph/9707243}}].

\bibitem{Isidori:2005xm}
G.~Isidori, F.~Mescia, and C.~Smith, {\it {Light-quark loops in $K
  \to\pi\nu\bar\nu$}},  {\em Nucl. Phys.} {\bf B718} (2005) 319--338,
  [\href{http://arxiv.org/abs/hep-ph/0503107}{{\tt hep-ph/0503107}}].

\bibitem{Mescia:2007kn}
F.~Mescia and C.~Smith, {\it {Improved estimates of rare K decay
  matrix-elements from $K_{\ell3}$ decays}},  {\em Phys. Rev.} {\bf D76} (2007)
  034017, [\href{http://arxiv.org/abs/0705.2025}{{\tt arXiv:0705.2025}}].

\bibitem{Buras:2004uu}
A.~J. Buras, F.~Schwab, and S.~Uhlig, {\it {Waiting for precise measurements of
  $K^{+} \to \pi^{+} \nu \bar{\nu}$ and $K_{L} \to \pi^0 \nu \bar{\nu}$}},
  {\em Rev. Mod. Phys.} {\bf 80} (2008) 965--1007,
  [\href{http://arxiv.org/abs/hep-ph/0405132}{{\tt hep-ph/0405132}}].

\bibitem{Komatsubara:2012pn}
T.~Komatsubara, {\it {Experiments with K-Meson Decays}},  {\em
  Prog.Part.Nucl.Phys.} {\bf 67} (2012) 995--1018,
  [\href{http://arxiv.org/abs/1203.6437}{{\tt arXiv:1203.6437}}].

\bibitem{Blanke:2013goa}
M.~Blanke, {\it {New Physics Signatures in Kaon Decays}},  {\em PoS} {\bf
  KAON13} (2013) 010, [\href{http://arxiv.org/abs/1305.5671}{{\tt
  arXiv:1305.5671}}].

\bibitem{Smith:2014mla}
C.~Smith, {\it {Rare K decays: Challenges and Perspectives}},
  \href{http://arxiv.org/abs/1409.6162}{{\tt arXiv:1409.6162}}.

\bibitem{Rinella:2014wfa}
G.~A. Rinella, R.~Aliberti, F.~Ambrosino, B.~Angelucci, A.~Antonelli, et~al.,
  {\it {Prospects for $K^+ \to \pi^+ \nu \bar{ \nu }$ at CERN in NA62}},
  \href{http://arxiv.org/abs/1411.0109}{{\tt arXiv:1411.0109}}.

\bibitem{Romano:2014xda}
A.~Romano, {\it {The $K^+ \rightarrow \pi^+ \nu \bar{\nu}$ decay in the NA62
  experiment at CERN}},  \href{http://arxiv.org/abs/1411.6546}{{\tt
  arXiv:1411.6546}}.

\bibitem{Shiomi:2014sfa}
{\bf KOTO} Collaboration, K.~Shiomi, {\it {$K^{0}_{L}\rightarrow \pi^{0} \nu
  \bar{\nu}$ at KOTO}},  \href{http://arxiv.org/abs/1411.4250}{{\tt
  arXiv:1411.4250}}.

\bibitem{Buras:2015qea}
A.~J. Buras, D.~Buttazzo, J.~Girrbach-Noe, and R.~Knegjens, {\it
  {$K^+\to\pi^+\nu\bar\nu$ and $K_L\to\pi^0\nu\bar\nu$ in the Standard Model:
  Status and Perspectives}},  \href{http://arxiv.org/abs/1503.02693}{{\tt
  arXiv:1503.02693}}.

\bibitem{Buras:2015yca}
A.~J. Buras, D.~Buttazzo, and R.~Knegjens, {\it {$K\to\pi\nu\bar\nu$ and
  $\epsilon'/\epsilon$ in Simplified New Physics Models}},
  \href{http://arxiv.org/abs/1507.08672}{{\tt arXiv:1507.08672}}.

\bibitem{Buras:2010wr}
A.~J. Buras, {\it {Minimal flavour violation and beyond: Towards a flavour code
  for short distance dynamics}},  {\em Acta Phys.Polon.} {\bf B41} (2010)
  2487--2561, [\href{http://arxiv.org/abs/1012.1447}{{\tt arXiv:1012.1447}}].

\bibitem{Buras:2012ts}
A.~J. Buras and J.~Girrbach, {\it {BSM models facing the recent LHCb data: A
  First look}},  {\em Acta Phys.Polon.} {\bf B43} (2012) 1427,
  [\href{http://arxiv.org/abs/1204.5064}{{\tt arXiv:1204.5064}}].

\bibitem{Buras:1997ij}
A.~J. Buras, A.~Romanino, and L.~Silvestrini, {\it $k \to\pi \nu\bar\nu$: A
  model independent analysis and supersymmetry},  {\em Nucl. Phys.} {\bf B520}
  (1998) 3--30, [\href{http://arxiv.org/abs/hep-ph/9712398}{{\tt
  hep-ph/9712398}}].

\bibitem{Colangelo:1998pm}
G.~Colangelo and G.~Isidori, {\it {Supersymmetric contributions to rare kaon
  decays: Beyond the single mass-insertion approximation}},  {\em JHEP} {\bf
  09} (1998) 009, [\href{http://arxiv.org/abs/hep-ph/9808487}{{\tt
  hep-ph/9808487}}].

\bibitem{Buras:1999da}
A.~J. Buras, G.~Colangelo, G.~Isidori, A.~Romanino, and L.~Silvestrini, {\it
  {Connections between $\epsilon'/\epsilon$ and rare kaon decays in
  supersymmetry}},  {\em Nucl. Phys.} {\bf B566} (2000) 3--32,
  [\href{http://arxiv.org/abs/hep-ph/9908371}{{\tt hep-ph/9908371}}].

\bibitem{Buras:2004qb}
A.~J. Buras, T.~Ewerth, S.~Jager, and J.~Rosiek, {\it {$K^+ \to \pi^+ \nu
  \bar\nu$ and $K_L \to \pi^0 \nu \bar\nu$ decays in the general MSSM}},  {\em
  Nucl. Phys.} {\bf B714} (2005) 103--136,
  [\href{http://arxiv.org/abs/hep-ph/0408142}{{\tt hep-ph/0408142}}].

\bibitem{Crivellin:2011sj}
A.~Crivellin, L.~Hofer, U.~Nierste, and D.~Scherer, {\it {Phenomenological
  consequences of radiative flavor violation in the MSSM}},  {\em Phys.Rev.}
  {\bf D84} (2011) 035030, [\href{http://arxiv.org/abs/1105.2818}{{\tt
  arXiv:1105.2818}}].

\bibitem{Buras:2006wk}
A.~J. Buras, A.~Poschenrieder, S.~Uhlig, and W.~A. Bardeen, {\it {Rare $K$ and
  $B$ decays in the Littlest Higgs model without T-parity}},  {\em JHEP} {\bf
  11} (2006) 062, [\href{http://arxiv.org/abs/hep-ph/0607189}{{\tt
  hep-ph/0607189}}].

\bibitem{Blanke:2009am}
M.~Blanke, A.~J. Buras, B.~Duling, S.~Recksiegel, and C.~Tarantino, {\it {FCNC
  Processes in the Littlest Higgs Model with T-Parity: a 2009 Look}},  {\em
  Acta Phys.Polon.} {\bf B41} (2010) 657--683,
  [\href{http://arxiv.org/abs/0906.5454}{{\tt arXiv:0906.5454}}].

\bibitem{Blanke:2015wba}
M.~Blanke, A.~J. Buras, and S.~Recksiegel, {\it {Quark flavour observables in
  the Littlest Higgs model with T-parity after LHC Run 1}},
  \href{http://arxiv.org/abs/1507.06316}{{\tt arXiv:1507.06316}}.

\bibitem{Blanke:2008yr}
M.~Blanke, A.~J. Buras, B.~Duling, K.~Gemmler, and S.~Gori, {\it {Rare K and B
  Decays in a Warped Extra Dimension with Custodial Protection}},  {\em JHEP}
  {\bf 03} (2009) 108, [\href{http://arxiv.org/abs/0812.3803}{{\tt
  arXiv:0812.3803}}].

\bibitem{Bauer:2009cf}
M.~Bauer, S.~Casagrande, U.~Haisch, and M.~Neubert, {\it {Flavor Physics in the
  Randall-Sundrum Model: II. Tree-Level Weak-Interaction Processes}},  {\em
  JHEP} {\bf 1009} (2010) 017, [\href{http://arxiv.org/abs/0912.1625}{{\tt
  arXiv:0912.1625}}].

\bibitem{Straub:2013zca}
D.~M. Straub, {\it {Anatomy of flavour-changing Z couplings in models with
  partial compositeness}},  {\em JHEP} {\bf 1308} (2013) 108,
  [\href{http://arxiv.org/abs/1302.4651}{{\tt arXiv:1302.4651}}].

\bibitem{Buras:2012dp}
A.~J. Buras, F.~De~Fazio, J.~Girrbach, and M.~V. Carlucci, {\it {The Anatomy of
  Quark Flavour Observables in 331 Models in the Flavour Precision Era}},  {\em
  JHEP} {\bf 1302} (2013) 023, [\href{http://arxiv.org/abs/1211.1237}{{\tt
  arXiv:1211.1237}}].

\bibitem{Buras:2014yna}
A.~J. Buras, F.~De~Fazio, and J.~Girrbach-Noe, {\it {Z-Z' mixing and Z-mediated
  FCNCs in $SU(3)_C \times SU(3)_L \times U(1)_X$ Models}},  {\em JHEP} {\bf
  1408} (2014) 039, [\href{http://arxiv.org/abs/1405.3850}{{\tt
  arXiv:1405.3850}}].

\bibitem{Buras:2012jb}
A.~J. Buras, F.~De~Fazio, and J.~Girrbach, {\it {The Anatomy of Z' and Z with
  Flavour Changing Neutral Currents in the Flavour Precision Era}},  {\em JHEP}
  {\bf 1302} (2013) 116, [\href{http://arxiv.org/abs/1211.1896}{{\tt
  arXiv:1211.1896}}].

\bibitem{Buras:2014sba}
A.~J. Buras, F.~De~Fazio, and J.~Girrbach, {\it {$\Delta I=1/2$ rule,
  $\varepsilon '/\varepsilon $ and $K\rightarrow \pi \nu \bar{\nu }$ in $Z'
  (Z)$ and $G' $ models with FCNC quark couplings}},  {\em Eur.Phys.J.} {\bf
  C74} (2014) 2950, [\href{http://arxiv.org/abs/1404.3824}{{\tt
  arXiv:1404.3824}}].

\bibitem{Blanke:2009pq}
M.~Blanke, {\it {Insights from the Interplay of $K\rightarrow \pi
  \nu\overline{\nu}$ and $\epsilon_K$ on the New Physics Flavour Structure}},
  {\em Acta Phys.Polon.} {\bf B41} (2010) 127,
  [\href{http://arxiv.org/abs/0904.2528}{{\tt arXiv:0904.2528}}].

\bibitem{Buras:1998ed}
A.~J. Buras and L.~Silvestrini, {\it Upper bounds on $k \to\pi\nu\bar\nu$ and
  $k_l\to\pi^0 e^+ e^-$ from $\varepsilon'/\varepsilon$ and $k_l \to\mu^+
  \mu^-$},  {\em Nucl. Phys.} {\bf B546} (1999) 299--314,
  [\href{http://arxiv.org/abs/hep-ph/9811471}{{\tt hep-ph/9811471}}].

\bibitem{Buras:2014zga}
A.~J. Buras, D.~Buttazzo, J.~Girrbach-Noe, and R.~Knegjens, {\it {Can we reach
  the Zeptouniverse with rare $K$ and $B_{s,d}$ decays?}},  {\em JHEP} {\bf
  1411} (2014) 121, [\href{http://arxiv.org/abs/1408.0728}{{\tt
  arXiv:1408.0728}}].

\bibitem{Batley:2002gn}
{\bf NA48} Collaboration, J.~Batley et~al., {\it {A Precision measurement of
  direct CP violation in the decay of neutral kaons into two pions}},  {\em
  Phys.Lett.} {\bf B544} (2002) 97--112,
  [\href{http://arxiv.org/abs/hep-ex/0208009}{{\tt hep-ex/0208009}}].

\bibitem{AlaviHarati:2002ye}
{\bf KTeV} Collaboration, A.~Alavi-Harati et~al., {\it {Measurements of direct
  CP violation, CPT symmetry, and other parameters in the neutral kaon
  system}},  {\em Phys.Rev.} {\bf D67} (2003) 012005,
  [\href{http://arxiv.org/abs/hep-ex/0208007}{{\tt hep-ex/0208007}}].

\bibitem{Abouzaid:2010ny}
{\bf KTeV} Collaboration, E.~Abouzaid et~al., {\it {Precise Measurements of
  Direct CP Violation, CPT Symmetry, and Other Parameters in the Neutral Kaon
  System}},  {\em Phys. Rev.} {\bf D83} (2011) 092001,
  [\href{http://arxiv.org/abs/1011.0127}{{\tt arXiv:1011.0127}}].

\bibitem{Bertolini:1998vd}
S.~Bertolini, M.~Fabbrichesi, and J.~O. Eeg, {\it {Theory of the CP violating
  parameter $\epsilon'/\epsilon$}},  {\em Rev.Mod.Phys.} {\bf 72} (2000)
  65--93, [\href{http://arxiv.org/abs/hep-ph/9802405}{{\tt hep-ph/9802405}}].

\bibitem{Buras:2003zz}
A.~J. Buras and M.~Jamin, {\it $\varepsilon'/\varepsilon$ at the nlo: 10 years
  later},  {\em JHEP} {\bf 01} (2004) 048,
  [\href{http://arxiv.org/abs/hep-ph/0306217}{{\tt hep-ph/0306217}}].

\bibitem{Pich:2004ee}
A.~Pich, {\it {$\varepsilon'/\varepsilon$ in the standard model: Theoretical
  update}},  \href{http://arxiv.org/abs/hep-ph/0410215}{{\tt hep-ph/0410215}}.

\bibitem{Cirigliano:2011ny}
V.~Cirigliano, G.~Ecker, H.~Neufeld, A.~Pich, and J.~Portoles, {\it {Kaon
  Decays in the Standard Model}},  {\em Rev.Mod.Phys.} {\bf 84} (2012) 399,
  [\href{http://arxiv.org/abs/1107.6001}{{\tt arXiv:1107.6001}}].

\bibitem{Bertolini:2012pu}
S.~Bertolini, J.~O. Eeg, A.~Maiezza, and F.~Nesti, {\it {New physics in
  $\epsilon'$ from gluomagnetic contributions and limits on Left-Right
  symmetry}},  {\em Phys.Rev.} {\bf D86} (2012) 095013,
  [\href{http://arxiv.org/abs/1206.0668}{{\tt arXiv:1206.0668}}].

\bibitem{Buras:1991jm}
A.~J. Buras, M.~Jamin, M.~E. Lautenbacher, and P.~H. Weisz, {\it Effective
  hamiltonians for $\delta s = 1$ and $\delta b = 1$ nonleptonic decays beyond
  the leading logarithmic approximation},  {\em Nucl. Phys.} {\bf B370} (1992)
  69--104.

\bibitem{Buras:1992tc}
A.~J. Buras, M.~Jamin, M.~E. Lautenbacher, and P.~H. Weisz, {\it Two loop
  anomalous dimension matrix for $\delta s = 1$ weak nonleptonic decays. 1.
  $\ord(alpha_s^2)$},  {\em Nucl. Phys.} {\bf B400} (1993) 37--74,
  [\href{http://arxiv.org/abs/hep-ph/9211304}{{\tt hep-ph/9211304}}].

\bibitem{Buras:1992zv}
A.~J. Buras, M.~Jamin, and M.~E. Lautenbacher, {\it Two loop anomalous
  dimension matrix for $\delta s = 1$ weak nonleptonic decays. 2.
  $\ord(\alpha\alpha_s)$},  {\em Nucl. Phys.} {\bf B400} (1993) 75--102,
  [\href{http://arxiv.org/abs/hep-ph/9211321}{{\tt hep-ph/9211321}}].

\bibitem{Ciuchini:1992tj}
M.~Ciuchini, E.~Franco, G.~Martinelli, and L.~Reina, {\it
  {$\varepsilon'/\varepsilon$ at the next-to-leading order in QCD and QED}},
  {\em Phys. Lett.} {\bf B301} (1993) 263--271,
  [\href{http://arxiv.org/abs/hep-ph/9212203}{{\tt hep-ph/9212203}}].

\bibitem{Buras:1993dy}
A.~J. Buras, M.~Jamin, and M.~E. Lautenbacher, {\it The anatomy of
  $\varepsilon'/ \varepsilon$ beyond leading logarithms with improved hadronic
  matrix elements},  {\em Nucl. Phys.} {\bf B408} (1993) 209--285,
  [\href{http://arxiv.org/abs/hep-ph/9303284}{{\tt hep-ph/9303284}}].

\bibitem{Ciuchini:1993vr}
M.~Ciuchini, E.~Franco, G.~Martinelli, and L.~Reina, {\it {The $\Delta S = 1$
  effective Hamiltonian including next-to-leading order QCD and QED
  corrections}},  {\em Nucl.Phys.} {\bf B415} (1994) 403--462,
  [\href{http://arxiv.org/abs/hep-ph/9304257}{{\tt hep-ph/9304257}}].

\bibitem{Buras:1999st}
A.~J. Buras, P.~Gambino, and U.~A. Haisch, {\it Electroweak penguin
  contributions to non-leptonic $\delta f = 1$ decays at nnlo},  {\em Nucl.
  Phys.} {\bf B570} (2000) 117--154,
  [\href{http://arxiv.org/abs/hep-ph/9911250}{{\tt hep-ph/9911250}}].

\bibitem{Brod:2010mj}
J.~Brod and M.~Gorbahn, {\it {$\epsilon_K$ at Next-to-Next-to-Leading Order:
  The Charm-Top-Quark Contribution}},  {\em Phys.Rev.} {\bf D82} (2010) 094026,
  [\href{http://arxiv.org/abs/1007.0684}{{\tt arXiv:1007.0684}}].

\bibitem{Buras:2015yba}
A.~J. Buras, M.~Gorbahn, S.~J{\"a}ger, and M.~Jamin, {\it {Improved anatomy of
  $\epe$ in the Standard Model}},  \href{http://arxiv.org/abs/1507.06345}{{\tt
  arXiv:1507.06345}}.

\bibitem{Cirigliano:2003nn}
V.~Cirigliano, A.~Pich, G.~Ecker, and H.~Neufeld, {\it {Isospin violation in
  $\epsilon^\prime$}},  {\em Phys.Rev.Lett.} {\bf 91} (2003) 162001,
  [\href{http://arxiv.org/abs/hep-ph/0307030}{{\tt hep-ph/0307030}}].

\bibitem{Cirigliano:2003gt}
V.~Cirigliano, G.~Ecker, H.~Neufeld, and A.~Pich, {\it {Isospin breaking in
  $K\to\pi\pi$ decays}},  {\em Eur. Phys. J.} {\bf C33} (2004) 369--396,
  [\href{http://arxiv.org/abs/hep-ph/0310351}{{\tt hep-ph/0310351}}].

\bibitem{Bai:2015nea}
Z.~Bai, T.~Blum, P.~Boyle, N.~Christ, J.~Frison, et~al., {\it {Standard-model
  prediction for direct CP violation in $K\to\pi\pi$ decay}},
  \href{http://arxiv.org/abs/1505.07863}{{\tt arXiv:1505.07863}}.

\bibitem{Blum:2015ywa}
T.~Blum, P.~Boyle, N.~Christ, J.~Frison, N.~Garron, et~al., {\it {$K
  \rightarrow \pi\pi$ $\Delta I=3/2$ decay amplitude in the continuum limit}},
  \href{http://arxiv.org/abs/1502.00263}{{\tt arXiv:1502.00263}}.

\bibitem{Buras:1985yx}
A.~J. Buras and J.-M. G\'erard, {\it {$1/N$ Expansion for Kaons}},  {\em
  Nucl.Phys.} {\bf B264} (1986) 371.

\bibitem{Bardeen:1986vp}
W.~A. Bardeen, A.~J. Buras, and J.-M. G\'erard, {\it {The $\Delta I = 1/2$ Rule
  in the Large $N$ Limit}},  {\em Phys.Lett.} {\bf B180} (1986) 133.

\bibitem{Buras:1987wc}
A.~J. Buras and J.~M. G\'erard, {\it {Isospin Breaking Contributions to
  $\epe$}},  {\em Phys. Lett.} {\bf B192} (1987) 156.

\bibitem{Buras:2015xba}
A.~J. Buras and J.-M. Gerard, {\it {Upper Bounds on $\epe$ Parameters $\bsi$
  and $\bei$ from Large N QCD and other News}},
  \href{http://arxiv.org/abs/1507.06326}{{\tt arXiv:1507.06326}}.

\bibitem{Buras:2014apa}
A.~J. Buras, {\it {$\Delta I=1/2$ Rule and $\hat B_K$ : 2014}},
  \href{http://arxiv.org/abs/1408.4820}{{\tt arXiv:1408.4820}}.

\bibitem{Bardeen:1986uz}
W.~A. Bardeen, A.~J. Buras, and J.-M. G\'erard, {\it {The $K\to\pi \pi$ Decays
  in the Large N Limit: Quark Evolution}},  {\em Nucl.Phys.} {\bf B293} (1987)
  787.

\bibitem{Bardeen:1986vz}
W.~A. Bardeen, A.~J. Buras, and J.-M. G\'erard, {\it {A Consistent Analysis of
  the $\Delta I = 1/2$ Rule for K Decays}},  {\em Phys.Lett.} {\bf B192} (1987)
  138.

\bibitem{Bardeen:1987vg}
W.~A. Bardeen, A.~J. Buras, and J.-M. G\'erard, {\it {The B Parameter Beyond
  the Leading Order of 1/N Expansion}},  {\em Phys.Lett.} {\bf B211} (1988)
  343.

\bibitem{Buras:2014maa}
A.~J. Buras, J.-M. G{\'e}rard, and W.~A. Bardeen, {\it {Large $N$ Approach to
  Kaon Decays and Mixing 28 Years Later: $\Delta I = 1/2$ Rule, $\hat B_K$ and
  $\Delta M_K$}},  {\em Eur.Phys.J.} {\bf C74} (2014), no.~5 2871,
  [\href{http://arxiv.org/abs/1401.1385}{{\tt arXiv:1401.1385}}].

\bibitem{Pallante:1999qf}
E.~Pallante and A.~Pich, {\it {Strong enhancement of $\varepsilon'/\varepsilon$
  through final state interactions}},  {\em Phys. Rev. Lett.} {\bf 84} (2000)
  2568--2571, [\href{http://arxiv.org/abs/hep-ph/9911233}{{\tt
  hep-ph/9911233}}].

\bibitem{Buras:2000kx}
A.~J. Buras et~al., {\it {Final state interactions and epsilon'/epsilon: A
  critical look}},  {\em Phys. Lett.} {\bf B480} (2000) 80--86,
  [\href{http://arxiv.org/abs/hep-ph/0002116}{{\tt hep-ph/0002116}}].

\bibitem{Buras:1994ec}
A.~J. Buras, M.~E. Lautenbacher, and G.~Ostermaier, {\it {Waiting for the top
  quark mass, $K^+ \to \pi^+ \nu\bar\nu$, $B_s^0 - \bar B_s^0$ mixing and CP
  asymmetries in $B$ decays}},  {\em Phys. Rev.} {\bf D50} (1994) 3433--3446,
  [\href{http://arxiv.org/abs/hep-ph/9403384}{{\tt hep-ph/9403384}}].

\bibitem{Buchalla:1994tr}
G.~Buchalla and A.~J. Buras, {\it $\sin2\beta$ from $k \to \pi \nu\bar\nu$},
  {\em Phys. Lett.} {\bf B333} (1994) 221--227,
  [\href{http://arxiv.org/abs/hep-ph/9405259}{{\tt hep-ph/9405259}}].

\bibitem{Buras:2001af}
A.~J. Buras and R.~Fleischer, {\it {Bounds on the unitarity triangle,
  $\sin2\beta$ and $K \to\pi \nu\bar\nu$ decays in models with minimal flavor
  violation}},  {\em Phys. Rev.} {\bf D64} (2001) 115010,
  [\href{http://arxiv.org/abs/hep-ph/0104238}{{\tt hep-ph/0104238}}].

\bibitem{Lehner:2015jga}
C.~Lehner, E.~Lunghi, and A.~Soni, {\it {Emerging lattice approach to the
  K-Unitarity Triangle}},  \href{http://arxiv.org/abs/1508.01801}{{\tt
  arXiv:1508.01801}}.

\bibitem{Gaillard:1974hs}
M.~Gaillard and B.~W. Lee, {\it {Rare Decay Modes of the K-Mesons in Gauge
  Theories}},  {\em Phys.Rev.} {\bf D10} (1974) 897.

\bibitem{Ellis:1977uk}
J.~R. Ellis, M.~K. Gaillard, D.~V. Nanopoulos, and S.~Rudaz, {\it {The
  Phenomenology of the Next Left-Handed Quarks}},  {\em Nucl. Phys.} {\bf B131}
  (1977) 285. [Erratum: Nucl. Phys.B132,541(1978)].

\bibitem{Buras:1992uf}
A.~J. Buras and M.~K. Harlander, {\it {A Top quark story: Quark mixing, CP
  violation and rare decays in the standard model}},  {\em Adv. Ser. Direct.
  High Energy Phys.} {\bf 10} (1992) 58--201.

\bibitem{Giudice:2015toa}
G.~F. Giudice, P.~Paradisi, and A.~Strumia, {\it {Indirect determinations of
  the top quark mass}},  \href{http://arxiv.org/abs/1508.05332}{{\tt
  arXiv:1508.05332}}.

\bibitem{Buras:1994rj}
A.~J. Buras, {\it {Precise determinations of the CKM matrix from CP asymmetries
  in B decays and $\klpn$ }},  {\em Phys. Lett.} {\bf B333} (1994) 476--483,
  [\href{http://arxiv.org/abs/hep-ph/9405368}{{\tt hep-ph/9405368}}].

\end{thebibliography}\endgroup
\end{document}